# NEW DIGITAL SIGNATURE PROTOCOL BASED ON ELLIPTIC CURVES

Ounasser Abid[1], Jaouad Ettanfouhi[2] and Omar Khadir[3]

[1,2,3]Laboratory of Mathematics, Cryptography and Mechanics, Department of Mathematics, Fstm , University of Hassan II Mohammedia-Casablanca, Morocco
[1]abidounasser@gmail.com
[2]ettanfouhi@gmail.com
[3]khadir@hotmail.com

## ABSTRACT

*In this work, a new digital signature based on elliptic curves is presented. We established its efficiency and security. The method, derived from a variant of ElGamal signature scheme, can be seen as a secure alternative protocol if known systems are completely broken.*

## 1. INTRODUCTION

A lot of practical algorithms for digital signature have been developed since the invention of the public key cryptography in the late 1970's [1,12,11]. Let us recall the principle of these methods. The signer Alice needs to possess two kinds of keys. The first one is private, must be kept secret and stored only locally. The second is public and must be largely diffused to be accessible to the other users. If Alice wishes to sign a particular message, a contract or a will $M$, she has to give the solutions of a hard mathematical equation depending of $M$ and of her public key. With the help of her private key, she is able to solve the problem. Bob, the judge or anybody, can verify that the solutions computed by Alice are valid. The algorithm is constructed in such a way that, for an adversary, without knowing Alice private key, it is computationally too hard to solve the considered equation. Nobody other than Alice can forge her signature and give the right answer, even who conceived the signature equation. In 1985, Koblitz [8] and Miller [10] proposed independently the use of elliptic curves in cryptography. They showed that there exist groups more complex than the traditional multiplicative group $((\mathbb{Z}/p\mathbb{Z})^*,.)$ where $\mathbb{Z}$ is the set of all integers and $p$ is a prime number. These structures are useful and practical in public key cryptography. Opponent's task became more complicated. In the same year, ElGamal [2] proposed a digital signature protocol based on the discrete logarithm problem. Since then, many similar schemes were elaborated and published [3, 4]. Among them, a new variant was presented by one of the authors [7] in 2010 and later, exploited for interesting research works connected to the networks privacy and anonymity [15,16].

Permanently, classical signature schemes are facing attacks more and more sophisticated. If these systems are completely broken, alternative protocols, previously designed, prepared and tested, would be useful.

In this article, we apply a variant of ElGamal scheme to build a new digital signature based on the elliptic curves. The efficiency of the method is discussed and its security analyzed.

The paper is organised as follows. In section 2, we review the basic ElGamal digital signature algorithm. A known variant [7] of this protocol is described in section 3. In section 4, we recall the standard definition of the group law defined on elliptic curves. Section 5 is devoted to our new digital signature method. We end with the conclusion in section 6.

In the sequel, we will respect ElGamal paper notations [2]. $\mathbb{N}, \mathbb{Z}$ are respectively the sets of integers and non-negative integers. For every positive integer $n$, we denote by $\mathbb{Z}/n\mathbb{Z}$ the finite ring of modular integers and by $(\mathbb{Z}/n\mathbb{Z})^*$ the multiplicative group of its invertible elements. Let $a, b, c$ be three integers. The great common divisor of $a$ and $b$ is denoted by $\gcd(a, b)$. We write $a \equiv b \,[c]$ if $c$ divides the difference $a - b$, and $a = b \bmod c$ if $a$ is the remainder in the division of $b$ by $c$. The bit-length of an integer $n$ is the number of bits in its binary representation.

We start by describing the original ElGamal signature scheme.

## 2. ELGAMAL SIGNATURE PROTOCOL

In this section we recall ElGamal signature scheme [2].

1. Alice chooses three numbers:

- p, a large prime integer.
- $\alpha$, a primitive root of the finite multiplicative group $(\mathbb{Z}/p\mathbb{Z})^*$.
- x, a random element in $\{1, 2, \ldots, p - 1\}$.

Then she computes $y = \alpha^x \bmod p$. Alice public keys are $(p, \alpha, y)$, and $x$ her private key.

2. To sign the message $m$, Alice needs to solve the equation :

$$\alpha^m \equiv y^r r^s \,[p] \qquad (1)$$

where $r, s$ are the unknown variables.
Alice fixes arbitrary $r$ to be $r = \alpha^k \bmod p$, where $k$ is chosen randomly and invertible modulo $p - 1$. Equation (1) is then equivalent to :

$$m \equiv xr + ks \,[p - 1] \qquad (2)$$

As Alice knows the secret key $x$, and as the integer $k$ is invertible modulo $p - 1$, she computes the other unknown variable $s$ by $s \equiv \frac{m-xr}{k}[p - 1]$.

3. Bob can verify the signature by checking that congruence (1) is valid for the variables $r$ and $s$ given by Alice. Observe that, in step 1, we need to know how to product prime integers. Generally, the running time for generating these numbers takes the most important part in the total running time. In [6], we obtained interesting experimental results, and concluded by suggesting some rapid procedures.

In the next section, we recall briefly a digital signature protocol that was conceived by one of the authors in 2011.

## 3. VARIANT OF ELGAMAL SIGNATURE PROTOCOL

We describe a variant of ElGamal signature protocol.
This variant [7] is based on the equation:

$$\alpha^t \equiv y^r r^s s^m \,[p] \qquad (3)$$

$r, s, t$ are three unknown parameters, and $(p, \alpha, y)$ are Alice public keys, $p$ is a large prime integer. $\alpha$ is a primitive root of the finite multiplicative group $(\mathbb{Z}/p\mathbb{Z})$. $y$ is calculated by $y = \alpha^x \bmod p$, $x$ is a random element in $\{1,2,\ldots,p-1\}$.

Let $m = h(M)$, where $h$ is a hash function, and $M$ the message to be signed by Alice.
To solve (3), Alice fixes arbitrary $r$ to be $r = \alpha^k \bmod p$, and $s$ to be $s = \alpha^l \bmod p$, where $k, l$ are chosen randomly in $\{1,2,\ldots,p-1\}$.
Equation (3) is then equivalent to

$$t \equiv rx + ks + lm \,[p-1] \qquad (4)$$

Since Alice knows the values of $r, s, k, l, m, x$ she can compute the third unknown variable $t$. Bob can verify the signature by checking the congruence (3).

This scheme has the advantage that it does not use the extended Euclidean algorithm for computing $k^{-1} \bmod (p-1)$.

We illustrate the technique by the example given by the author of this variant [7].

**Example 3.1.** Let $(p, \alpha, y)$ be Alice public keys where $p = 509$, $\alpha = 2$ and $y = 482$. We emphasise that we are not sure if using a small value of $\alpha$ does not weaken the system. The private key is $x = 281$. Suppose that Alice wants to produce a signature for the message M for which $m \equiv h(M) \equiv 432\,[508]$ with the two random exponents $k = 208$ and $l = 386$. She computes $r \equiv \alpha^k \equiv 2^{208} \equiv 332\,[p]$, $s \equiv \alpha^l \equiv 2^{286} \equiv 39\,[p]$ and $t \equiv rx + ks + lm \equiv 440\,[p-1]$.

Bob or anyone can verify the relation $\alpha^t \equiv y^r r^s s^m\,[p]$. Indeed, we find that $\alpha^t \equiv 436\,[p]$ and $y^r r^s s^m \equiv 436\,[p]$. Notice here that $k$ and $l$ are even integers unlike in ElGamal protocol where the exponent $k$ is always odd since it must be relatively prime with $(p-1)$.

## 4. GROUP LAW ON ELLIPTIC CURVES

In this section we recall the additive operation on points belonging to an elliptic curve $(E)$. For more details, we refer the reader to [8,9,10,11,12].

Suppose that the equation of $(E)$ is

$$y^2 \equiv x^3 + ax + b\,[p], \qquad (5)$$

where $p$ is a prime integer and $a, b \in \{1,2,\ldots,p-1\}$.

Let $P(x_1, y_1)$ and $Q(x_2, y_2)$ two points on the curve $(E)$ and $\mathcal{O}$ an imaginary point at infinity.
1. If $x_1 \neq x_2$, then

15

$$\begin{cases} x_3 \equiv \lambda^2 - x_1 - x_2 \ [p], \\ y_3 \equiv \lambda(x_1 - x_3) - y_1 \ [p] \end{cases} \quad \lambda \equiv \frac{y_2 - y_1}{x_2 - x_1} \ [p]$$

2. If $x_1 = x_2$ and $y_1 = y_2$ then $R = \mathcal{O}$.
3. If $P = Q$ and $y_1 = 0$ then $R = \mathcal{O}$.
4. If $P = Q$ and $y_1 \neq 0$ then

$$\begin{cases} x_3 \equiv \lambda^2 - 2x_1 \ [p], \\ y_3 \equiv \lambda(x_1 - x_3) - y_1 \ [p] \end{cases} \quad \lambda \equiv \frac{3x_1^2 + a}{2y_1} \ [p]$$

With this additive law, the elliptic curve becomes a finite Abelian group. Its structure seems to be more complex than the traditional multiplicative group $\left(\frac{\mathbb{Z}}{p\mathbb{Z}}\right)^*$.
We move to the next section where we describe our method.

## 5. OUR SIGNATURE PROTOCOL BASED ON ELLIPTIC CURVES

In this section we present our signature scheme. For elliptic curves digital signature algorithm (ECDSA) based partially on the ElGamal classical protocol, see [14, p.297] or [8, p.134]. Our method has some advantages and can be seen as an alternative if known systems are completely broken. Unlike what happens with other algorithms, we don't need to compute any modular inverse.

### 5.1. Description of our protocol :

Let $p$ be a prime integer and $(E)$ the elliptic curve defined by

$$y^2 = x^3 + ax + b \ [p] \qquad (6)$$

From the last section, we know that the points of the elliptic curve, with the particular point at infinity $\mathcal{O}$, form an additive Abelian group $(E, +)$.

Let G be an element of the curve $E$ whose order is a large prime integer $q$. We put B = G where is taken randomly in $\{1,2,\ldots,q-1\}$ as Alice private key.

For a message $M$, we compute $m = h(M)$ where $h$ is a secure hash function, like SHA1[9,14]. We suppose that $m < q$.

Alice public keys are $(p, a, b, G, B)$. We propose the following new protocol:
If Alice wants to sign the message $M$, she has to give the solutions of the equation:

$$tG = sR + rS + mB \qquad (7)$$

Where $R(r, r'), S(s, s')$ are two unknown points belonging to the curve $E$. Points $R, S$ and integer $t$ are to be determined by Alice. Her signature is $(R, S, t)$ is formed by two points and a natural integer $t < q$.

To solve the equation (7) Alice sets $R = kG$, $S = lG$ where $k$ and $l$ are chosen randomly in $\{1, 2, \ldots, q - 1\}$. Equation (7) is equivalent to: $tG = skG + rlG + m\,G$ and as $q$ is the order of the point $G$, we obtain:

$$t \equiv sk + rl + m\alpha \; [q] \tag{8}$$

Knowing $\alpha, r, s, k$ and $l$ Alice is able to compute the last unknown variable $t$. So she has all the solutions of relation (7).

Notice that there exist efficient algorithms for the elliptic curve scalar multiplication, see for instance [5].

In another hand, Bob can verify the validity of the solutions by replacing $R, S$ and $t$ in the original relation.

Observe that in our scheme, unlike ECDSA[8,14], we don't need to compute any modular inverse.

Let us illustrate the method by the following example.

**Example 5.1.** Let $(E)$ be the elliptic curve defined by $y^2 \equiv x^3 + 6x + 2 \; [757]$.

We find that the cardinality of $(E)$ is $n = 791$. The point $G = (529, 566)$ has as order the prime integer $q = 113$.

Assume that Alice private key is $\alpha = 78$ so $B = \alpha G = (319, 629)$.
Alice public keys are therefore $(p, a, b, G, B)$.
Suppose that the hashed of the message $M$ is $m \equiv h(M) \equiv 56 \; [q]$ and that Alice wants to sign the message $M$.

Let us admit that Alice chooses the random exponents $k = 81$ and $l = 63$.
She calculates $R = kG = (248, 195)$ and $S = lG = (157, 326)$.
By formula (8): $t \equiv sk + rl + m\alpha \equiv 52 \; [q]$.
So the signature is $(R, S, t) = (248, 195, 157, 326, 52)$.
We can check that $tG = (555, 156)$, $sR = (555, 601)$, $rS = (292, 266)$, $mB = (26, 319)$, and therefore $tG = sR + rS + mB$.

**Remark 5.1:** Alice can sign two messages with the same secret couple $(k, l)$ without risking to reveal her private key $\alpha$. Indeed, let $(R, S, t_1)$ and $(R, S, t_2)$ be the signature of two different messages $M_1$ and $M_2$ associated to the secret couple $(k, l)$. We have

$$\begin{cases} t_1 \equiv sk + rl + m_1\alpha \; [q] \\ t_2 \equiv sk + rl + m_2\alpha \; [q] \end{cases}$$

where $m_1 = h(M_1)$ and $m_2 = h(M_2)$.

We have two modular equations and three unknown variables $k, l$ and $\alpha$. It seems that it is not an easy task to retrieve secret parameters $k$ and $l$. Notice that parts of papers [15] and [16] are based on a similar remark figuring in [7].

### 5.2. Security analysis :

Now that we presented the protocol, we start discussing some eventual attacks. Assume that Oscar is Alice opponent.

**Attack 1 :** Suppose that Oscar knows Alice signature for a message $M$, and he tries to find Alice private key $\alpha$. As formula (7) is equivalent to $t \equiv sk + rl + m\alpha\ [q]$, Oscar cannot compute $\alpha$ since he ignores the values of the parameters $k$ and $l$.

**Attack 2 :** Suppose that Oscar tries to forge Alice signature by fixing arbitrary two parameters and looking for the third:

**(1)** If Oscar fixes $R$ and $S$ and aims to compute $t$ in equation (7), he will be confronted to the elliptic curves discrete logarithm problem.

**(2)** If Oscar fixes $R$ and $t$ or $S$ and t, he will have from formula (7), $sR + rS = tG - mB$ and there is no known way to solve this type of equations.

**Attack 3 :** Suppose that Oscar has obtained different valid signatures for $z$ messages $M_i$, $i \in \{1,2,\ldots,z-1\}$, $z \in \mathbb{N}$. If we put $m_i = h(M_i)$, he will get a system of $z$ modular equations:

$$(S) \begin{cases} t \equiv s_1 k_1 + r_1 l_1 + m_1 \alpha\ [q] \\ t \equiv s_2 k_2 + r_2 l_2 + m_2 \alpha\ [q] \\ \quad \vdots \\ t \equiv s_z k_z + r_z l_z + m_z \alpha\ [q] \end{cases}$$

Since system $(S)$ contains $2z + 1$ unknown variables $\alpha, k_i, l_i$, $i \in \{1,2,\ldots,z-1\}$, Oscar will find many solutions. He cannot know the correct one due to the uniqueness of Alice private key .

### 5.3. Complexity of our method :

As in [4], let $T_{ECmult}$, $T_{MOmult}$, $T_h$, be respectively the time to perform a multiple point in an elliptic curve, a modular multiplication and a hash function computation of a message $M$. We ignore the time required for modular additions, substractions, comparisons and make the conversion $T_{ECmult} = 240 T_{MOmult}$.

The signer Alice needs to perform two multiple points, three modular multiplications and one hash function computation. So the global required time is : $T_1 = 2T_{ECmult} + 3T_{MOmult} + T_h = 483 T_{MOmult} + T_h$.

The verifier Bob needs to perform four multiple points in an elliptic curve and one hash function computation. So the global required time is : $T_2 = 4T_{ECmult} + T_h = 960 T_{MOmult} + T_h$
The cost of communication, without $M$, is $12|p|$, since to sign, Alice transmits $(p, a, b, G, B)$ and $(R, S, t)$; |p| denoting the bit-length of the integer $p$.

## 6. CONCLUSION

In this work we presented a new digital signature. A variant of ElGamal signature scheme was applied to the elliptic curves. We also analyzed its security and efficiency. The method can be seen as a practical alternative system if known protocols are completely broken.

## REFERENCES

[1] W. Diffie and M. E. Hellman, New directions in cryptography, IEEE Transactions on Information Theory, vol. IT-22, (1976), pp. 644-654.
[2] T. ElGamal, A public key cryptosystem and a signature scheme based on discrete logarithm problem, IEEE Trans. Info. Theory, IT-31, (1985), pp. 469-472.
[3] P. Horster, M. Michels, H. Petersen, Generalized ElGamal signature schemes for one message block, Technical Report, TR-94-3, 1994.
[4] E. S. Ismail, N. M. F. Tahat and R. R. Ahmad, A new digital signature scheme based on factoring and discrete logarithms, J. of Mathematics and Statistics (4): (2008), pp. 222-225.
[5] V. S. Iyengar, A novel elliptic curve scalar multiplication algorithms for faster and safer public-key cryptosystems, Int. J. of Cryptography and Information security, Vol. 2, n° 3, (2012), pp. 57-66.
[6] O. Khadir, L. Szalay, Experimental results on probable primality, Acta Univ. Sapien-tiae, Math. 1, no. 2, (2009), pp. 161-168. Available at http://www.emis.de/journals/AUSM/C1-2/math2-6.pdf
[7] O. Khadir, New variant of ElGamal signature scheme, Int. J. Contemp. Math. Sciences, Vol. 5, no. 34, (2010), pp. 1653-1662. Available at http://www.m-hikari.com/ijcms-2010/33-36-2010/khadirIJCMS33-36-2010.pdf
[8] N. Koblitz, Elliptic curve cryptosystem, Math. Comp. 48 (1987), pp. 203-209.
[9] A. J. Menezes, P. C. van Oorschot and S. A. Vanstone, Handbook of applied cryptography, CRC Press, Boca Raton, Florida, 1997. Available at http://www.cacr.math.uwaterloo.ca/hac/
[10] V. Miller, Uses of elliptic curves in cryptography , in: H.C. Williams (Ed.), Advances in Cryptology: Proceedings of Crypto'85, Lecture Notes in Computer Science, Vol. 218, Springer, Berlin, 1985, pp. 417-426.
[11] M. O. Rabin, Digitalized signatures and public key functions as intractable as factoring, MIT/LCS/TR, Vol. 212, (1979).
[12] R. Rivest, A. Shamir and L. Adeleman, A method for obtaining digital signatures and public key cryptosystems, Communication of the ACM, Vol. no 21, (1978), pp. 120-126.
[13] J. H. Silverman, The arithmetic of elliptic, Springer Verlag, Berlin, 1986.
[14] D. R. Stinson, Cryptography, theory and practice, Third Edition, Chapman & Hall/CRC, 2006.
[15] V. J. Vazram, V.V. Kumari and J.V.R. Murthy, Privacy in mobile ad hoc networks, Advances in digital image processing and information technology, CCIS, Vol. 205, Springer, pp. 336-345, 2011.
[16] V. J. Vazram, V.V. Kumari and J.V.R. Murthy, Anonymity and security in mobile ad hoc networks, Distributed Computing and Internet Technology, 8th International Conference, LNCS 7154, Springer, pp. 71-82, 2012.

## Authors

### Short biography

**Ounasser Abid** holds an engineer degree in Computer Science from the University of Hassan II Mohammedia (2011). Member of the laboratory of Mathematics, Cryptography and Mechanics, he is preparing a thesis in public key cryptography.

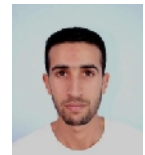

Jaouad Ettanfouhi holds an engineer degree in Computer Science from the University of Hassan II Mohammedia (2011). Member of the laboratory of Mathematics, Cryptography and Mechanics, he is preparing a thesis in public key cryptography.

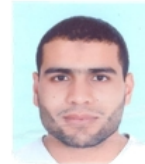

Dr Omar Khadir received his Ph.D. degree in Computer Science from the University of Rouen, France (1994). Co-founder of the Laboratory of Mathematics,

Cryptography and Mechanics at the University of Hassan II Mohammedia, Morocco, where he is a professor in the Department of Mathematics. He teaches cryptography for graduate students preparing a degree in computer science.

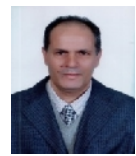

His current research interests include public key cryptography, digital signature, primality, factorisation of large integers and more generally, all subjects connected to the information technology.